\documentclass[pre,aps,twocolumn,superscriptaddress]{revtex4}
\usepackage{graphicx}
\usepackage{bm}

\usepackage[latin1]{inputenc}
\newcommand{\beq}{\begin{equation}}
\newcommand{\eeq}{\end{equation}}
\newcommand{\bea}{\begin{eqnarray}}
\newcommand{\eea}{\end{eqnarray}}
\newcommand{\K}{K_{_{\rm I}}(v,L)}
\newcommand{\R}{K^2_{_{\rm I}}(v,L)}
\begin{document}
\title{Velocity Fluctuations in Dynamical Fracture: the Role of Microcracks \\ Version of \today}
\author{Eran Bouchbinder}
\affiliation{Dept. of Chemical Physics, The Weizmann Institute
of Science, Rehovot 76100, Israel}
\author{David Kessler}
\affiliation{Dept. of Physics, Bar--Ilan University, Ramat Gan 52900, Israel}
\author{Itamar Procaccia}
\affiliation{Dept. of Chemical Physics, The Weizmann Institute
of Science, Rehovot 76100, Israel}
\begin{abstract}

We address the velocity fluctuations of fastly moving cracks in
stressed materials. One possible mechanism for such fluctuations
is the interaction of the main crack with micro cracks
(irrespective whether these are existing material defects or they
form during the crack evolution).  We analyze carefully the
dynamics (in 2 space dimensions) of one macro and one micro crack,
and demonstrate that their interaction results in a {\em large}
and {\em rapid} velocity fluctuation, in qualitative
correspondence with typical velocity fluctuations observed in
experiments.  In developing the theory of the dynamical
interaction we invoke an approximation that affords a reduction in
mathematical complexity to a simple set of ordinary differential
equations for the positions of the cracks tips; we propose that
this kind of approximation has a range of usefulness that exceeds
the present context.

\end{abstract}
\maketitle

\section{Introduction}

Classical linear elasticity fracture mechanics provides clear cut
predictions for the dynamical evolution of cracks in stressed
materials. Under pure mode I loading a crack is expected to remain
straight, and to exhibit a tip velocity that increases
monotonically towards the Rayleigh wave speed $c_R$. Reality shows
that this is but a pipe dream. When the crack velocity exceeds a
finite fraction of $c_R$ the velocity of typical cracks exhibits
wild fluctuations, the crack surfaces lose their smoothness and
the mean velocity never asymptotes towards $c_R$. The fundamental
understanding of the discrepancy between the prediction of the
classical theory and experiments remains an open problem of
considerable interest and importance.

A number of studies
\cite{84R-CK_1,84R-CK_2,95SF,96SF,97R-CY,98R,99SF,99FM} point
towards a close correspondence between the onset of velocity
fluctuations and the appearance of secondary damage like micro
cracks (appearing ahead of the tip), microscopic side branches
etc. Conical markings which are observed on crack surfaces offer a
good indication that micro cracks exist before the arrival of the
crack, although it is not determined whether the former stem from
material imperfections or from stress instabilities.  The fact
that the density of conical markings increases during the crack
evolution \cite{97R-CY} suggests that the level of stress is
responsible in some way for the activation of the micro cracks.
The aim of this paper is to explore the connection between
velocity fluctuations and the putative existence of micro cracks
ahead of the crack tip. To this aim will study the dynamical
interaction between a macro crack and a micro crack and focus on
the velocity of the tip of former under the influence of the
latter.

To actually solve exactly the dynamical equations for the displacement field with boundary
condition on both macro and micro cracks up to coalescence is a very taxing quest. Building upon
experience in the field we will propose here an approximate methodology that will allow us writing
down {\em ordinary} differential equations for the positions of the tips of both macro and micro cracks.
While sensible, the approximate methodology is not established in a controlled fashion, requiring
therefore simulational support. Indeed, we will offer in this paper lattice simulations to back
the analytic considerations. We will show that the correspondence is excellent.

In Sect. \ref{setup} we introduce the problem at hand, being an
infinite 2-dimensional stressed material with one macro crack and
one colinear micro crack of length $\ell$. In Sect. \ref{Approx} we describe the
approximate method of solution, motivating it by the exactly soluble
cases of straight and bifurcating cracks. The Section culminates with approximate
equations of motion for the tips of the macro and micro cracks. In Sect. \ref{solution} we describe
the solution of the model problem, stressing the velocity of the tip of the macro crack.
We show that the net result of the interaction is a rapid and large up and down fluctuation
in this velocity, in correspondence with the observed fluctuations in dynamical crack propagation.
Sect. \ref{simul} provides a simulational support to the approximate theory; by performing
lattice simulations we study the same model problem and compare the results. The close
correspondence between approximate theory and simulations lends support to the former.
Sect. \ref{summary} offers a summary and conclusions.

\section{The problem}
\label{setup}

\begin{figure}[here]
\centerline{\includegraphics[width=.48\textwidth]{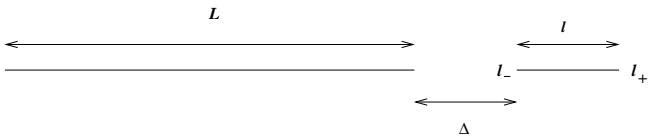}}
\caption{The geometric configuration of the model problem.  The
macro crack and the micro crack extend along the intervals
$[-L,0]$ and $[\ell_-,\ell_+]$ respectively with $L/\ell \gg 1$.}
\label{Sketch}
\end{figure}

The problem that we want to consider is sketched in Fig.
\ref{Sketch}. We consider a macro crack and a micro crack that
at a given time extend along the intervals $[-L,0]$ and $[\ell_-,\ell_+]$
respectively. The distance between them is given by $\Delta$. The
length of the micro crack is $\ell\equiv \ell_+- \ell_-$. We
expect on physical grounds that the micro cracks in typical
materials are at most of the size of the process zone, and
therefore we always consider the limit $\ell/L\to 0$.\\
The aim of the calculation is to determine the simultaneous motion
of the three crack tips (the macro crack tip, the inner and outer
tips of the micro crack) as a function of time. In full generality
this entails the general solution of the field equations for an
arbitrary motion, specifically the determination of the {\em
dynamic} stress intensity factors at the cracks tips and then to
apply a fracture criterion to obtain the actual dynamics. We
cannot offer an exact solution to this problem. Instead, we will
introduce an approximate method that provides analytic
insight to the problem.
\section{Approximate method of solution}
\label{Approx}

To motivate our approximate methodology we will recall some exact classical results obtained for
ideal (mode I) straight dynamical cracks and more recent results pertaining to (mode III) bifurcating cracks.
\subsection{Motivation I: Ideal Straight Cracks}
\label{straight}

\begin{figure}[here]
\centerline{\includegraphics[width=.47\textwidth]{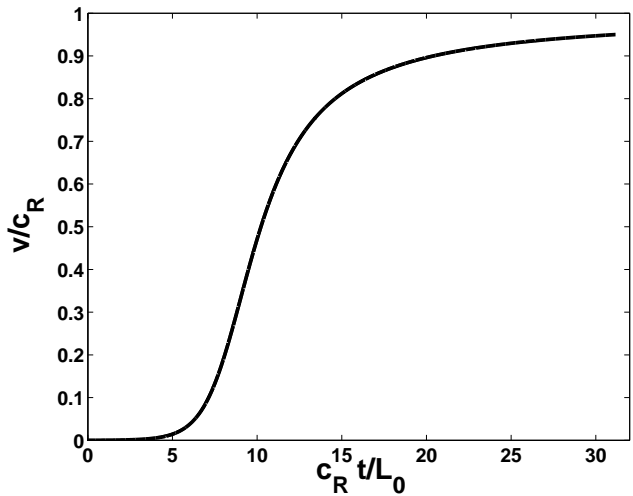}}
\caption{The predicted velocity increase as a function of
normalized time (where $L_0$ is the crack length at initiation)
for a mode I crack under uniform constant load at infinity, Eq.
(\ref{BasicEq}).} \label{smooth}
\end{figure}

As said above, linear elasticity fracture mechanics provides exact
solutions for straight cracks under mode I loading. The stress
field $\sigma_{ij}(r,\theta,t)$, measured in polar coordinates
relative to the tip, is expected to have a universal form in the
vicinity of the crack tip \cite{98F},
\begin{equation}
\sigma_{ij}(r,\theta,t) = K_{_{\rm I}}(v(t),L(t)) \frac{\Sigma^{^{\rm I}}_{ij}(\theta,v(t))}{\sqrt{2\pi r}} \ . \label{sigma}
\end{equation}
Here $v(t)$ and $L(t)$ are the time-dependent crack velocity and
length respectively, $\Sigma^{^{\rm I}}_{ij}(\theta,v)$ are known
universal functions \cite{98F}, and $\K$ is the ``stress intensity factor"
which is predicted to depend on the instantaneous crack velocity
and length {\em only}. For notational simplicity we drop the $t$
dependence. At each moment in time the velocity is expected (for
plane stress conditions) to be determined by the energy balance
equation \cite{98F}
\begin{equation}
\Gamma(v) = \frac{1}{E}A_{_{\rm I}}(v) \R \ .
\label{BE}
\end{equation}
The LHS here is the fracture energy and the RHS is the energy
release rate into the crack tip region, resulting from a path
integral over the total energy flux. $E$ is Young's modulus and
$A_{_{\rm I}}(v)$ is a mode I universal function. The exact result
that we refer to is the decomposition of the dynamic stress
intensity factor $\K$ for a semi-infinite crack under time
independent loading, in the form \cite{98F,72F,75Kos}
\begin{equation}
\K =  k_{_{\rm I}}(v)~K^s_{_{\rm I}}(L) \ , \label{funda}
\end{equation}
where $k_{I}(v)$ is a universal function of $v$ and $K^s_{_{\rm I}}(L)$ is the stress intensity factor of
a {\em static} crack of length $L$ under the same loading (when $L$ is large enough to
be considered as semi-infinite). This important result is the basis of the classical theory
of straight crack motion. The calculation of the static stress intensity factor is a much
easier task than the evaluation of its dynamical counterpart, since it requires solutions
of bi-Laplace equations with boundary conditions.
Rewriting Eq. (\ref{BE}) in the light of this result one obtains,
\begin{equation}
\Gamma(v) = \frac{1}{E}A_{_{\rm I}}(v) [k_{_{\rm I}}(v)~K^s_{_{\rm
I}}(L)]^2. \label{BE1}
\end{equation}
A further serendipitous simplification arises from numerical
evaluations of the combination $A_{_{\rm I}}(v) k^2_{_{\rm
I}}(v)$, showing that it is well approximated by \cite{98F}
\begin{equation}
A_{_{\rm I}}(v) k^2_{_{\rm I}}(v) \approx 1-v/c_R.
\end{equation}
This approximation leads to an ordinary differential equation for
the crack length. If one asserts that the fracture energy $\Gamma$
is $v$-independent this differential equation becomes explicit,
\begin{equation}
\frac{dL(t)}{dt} \approx c_R\left[1-\frac{E\Gamma}{[K^s_{_{\rm I}}\big(L(t)\big)]^2}\right].
\label{BasicEq}
\end{equation}
It should be stressed again that the decomposition property of the
dynamic stress intensity factor (Eq. (\ref{funda})) is essential
in deriving this basic equation. This equation predicts a
monotonic increase in the tip velocity asymptoting towards $c_R$.
There is no crack motion as long as the stress intensity factor does
not exceed a material dependent threshold
\begin{equation}
K^s_{_{\rm I}}\big(L(t)\big) < \sqrt{E\Gamma}\quad\text{no crack motion} \ . \label{crit}
\end{equation}
For the sake of illustration we show in Fig. \ref{smooth} the
solution of this equation for the simple case $K^s_{_{\rm
I}}\big(L(t)\big)=\sigma^\infty\sqrt{\pi L(t)/2}$, where
$\sigma^\infty$ is the load at infinity.

\subsection{Motivation II: Bifurcating Cracks}
\label{bifurcate}

The bifurcation of fast cracks is observed in many experiments,
and understanding it theoretically is a problem of some importance
in the theory of fracture. One interesting problem that was
addressed recently \cite{03A-B} in this context is the determination
of the stress intensity factors at the tips of symmetrically
branched cracks in terms of the stress intensity factor prior to
branching. Ref. \cite{03A-B} presented a solution to this problem
for mode III (anti-plane) conditions. In this solution the stress
intensity factor $K'$ at the tips of two symmetric branches
emerging from a macro crack at a velocity $v'$ and creating an
angle $\lambda\pi$ relative to the macro crack line, was given in
the form
\begin{equation}
K' = \sqrt{1-v'/c_s}~H_{33}(\lambda,v'/c_s)~ K_0 \ .
\label{Adda-Bedia}
\end{equation}
Here $\sqrt{1-v'/c_s}$ is the mode III universal function, whose
mode I counterpart is $k_{_{\rm I}}(v)$, $c_s$ is the shear wave
speed, $K_0$ is the stress intensity factor of the macro crack
prior to branching and $H_{33}(\lambda,v'/c_s)$ carries the
information regarding the dynamic interaction. Note that the macro
crack velocity $v$ is absent here as in the branching scenario
adopted in \cite{03A-B} the macro crack stops suddenly before
branching and a static stress distribution is established behind a
wave front travelling at the characteristic wave speed $c_s$. If
the decomposition in Eq. (\ref{funda}) is of some generality then
we expect the main interaction effect to be contained in the {\em
static} stress intensity factors for the bifurcated configuration
given by $H_{33}(\lambda,0) K_0$. Indeed, Fig. 4 in ref.
\cite{03A-B} shows that ratio
\begin{equation}
 \frac{H_{33}(\lambda,v'/c_s)}{H_{33}(\lambda,0)}
\end{equation}
is very close to unity (up to $\pm5\%$) for all the values of $\lambda$
and $v'/c_s$. Therefore, we conclude that even for this complex
configuration the dynamic stress intensity factor admits an
approximate decomposition in the form of a product of its static
counterpart and a universal function of the local crack tip
velocity,
\begin{equation}
K' \approx \sqrt{1-v'/c_s}~H_{33}(\lambda,0)~ K_0 \ . \label{adda}
\end{equation}
This result suggests that a very good approximation for the {\em
dynamic} stress intensity factor can be obtained by calculating
the {\em static} stress intensity factor for the same
instantaneous configuration and a knowledge of a universal
velocity function characteristic of the local symmetry conditions
at the crack tip.

\subsection{The Decomposition Approximation for our Problem}

For advancing the problem posed in this paper we need to consider the stress field in the vicinity of three
crack tips. In the vicinity of the tip of the macro crack we write (in the local polar
coordinates $(r_{_{\rm M}}, \theta_{_{\rm M}})$ around that tip)
\begin{equation}
\sigma_{ij}(r_{_{\rm M}}, \theta_{_{\rm M}},t) = K_{_{\rm M}} \frac{\Sigma^{^{\rm
I}}_{ij}(\theta_{_{\rm M}},v_{_{\rm M}}(t))}{\sqrt{2\pi r}}
\ ,
\label{sigmaM}
\end{equation}
where $K_{_{\rm M}}$ is the stress intensity factor that in
principle depends on the positions and velocities of all the tips
[i.e. $L(t),\ell_\pm(t), v_{_{\rm M}}(t),v_\pm(t)$] and maybe
other derivatives. Near the tips of the micro crack we write
similarly (in local polar coordinates $(r_\pm,\theta_\pm)$ around
each tip)
\begin{equation}
\sigma_{ij}(r_\pm,\theta_\pm,t) = K_{\pm} \frac{\Sigma^{^{\rm I}}_{ij}(\theta_\pm,v_\pm(t))}{\sqrt{2\pi r}}
\ ,
\label{sigmapm}
\end{equation}
where $K_\pm$ are again the stress intensity factors that depend
on all the time dependent functions $L(t),\ell_\pm(t), v_{_{\rm
M}}(t),v_\pm(t)$ and maybe other derivatives.

Our basic approximation is now motivated by the two examples Eqs.
(\ref{funda}) and (\ref{adda}); we assume that the dynamic stress
intensity factors can be decomposed according to
\begin{eqnarray}
K_{_{\rm M}}&\approx&k_{_{\rm I}}(v_{_{\rm M}}) K^s_{_{\rm M}} \Big(L(t),\ell_+(t),\ell_-(t)\Big)\nonumber\\
K_+&\approx&k_{_{\rm I}}(v_+) K^s_+ \Big(L(t),\ell_+(t),\ell_-(t)\Big) \nonumber\\
K_-&\approx&k_{_{\rm I}}(v_-) K^s_+
\Big(L(t),\ell_+(t),\ell_-(t)\Big)\ . \label{approx}
\end{eqnarray}
Here the universal function $k_{_{\rm I}}(v)$ is the {\em same}
function appearing in Eq. (\ref{funda}) and all the stress
intensity factors with superscript $s$ refer to the solution of
the {\em static} problem with a frozen geometry which is given by
the crack tip positions $L(t), \ell_\pm (t)$. On physical grounds
we expect this approximation to be good when $\ell/L\to 0$, and to
lose its validity as this ratio increases. The numerical
simulations presented in Sect. \ref{simul} lend a strong support
to this expectation.

The advantage of this approximation is that it leads to ordinary differential equations for the
tip positions in much the same way that Eq. (\ref{BasicEq}) followed from Eq. (\ref{funda}),
\begin{eqnarray}
\frac{dL(t)}{dt} &\approx& c_R\left[1-\frac{E\Gamma}{[K^s_{_{\rm M}}\big(L(t),\ell_+(t),\ell_-(t)\big)]^2}\right]\nonumber\\
\frac{d\ell_-(t)}{dt} &\approx& c_R\left[1-\frac{E\Gamma}{[K^s_-\big(L(t),\ell_+(t),\ell_-(t)\big)]^2}\right]\nonumber\\
\frac{d\ell_+(t)}{dt} &\approx&
c_R\left[1-\frac{E\Gamma}{[K^s_+\big(L(t),\ell_+(t),\ell_-(t)\big)]^2}\right].
\label{DynSys}
\end{eqnarray}\\
We turn now to the analysis of this set of equations and their consequences.
\section{Solution of the model}
\label{solution}
\subsection{The static problem}

A prerequisite to the solution of the set of equations (\ref{DynSys}) is the calculation
of the static stress intensity factors for a general configuration of two colinear cracks.
We employ the available solution for two colinear cracks
consisting of segments $a<x<b$ and $c<x<d$ with $b<c$ under a remote mode I
loading $\sigma^{\infty}$. The $\sigma_{yy}$ component of the stress tensor
along the cracks line, outside the cracks, is given by \cite{99B}
\begin{eqnarray}
\sigma_{yy}(x,0)&=&\frac{\sigma^{\infty}}{2G(x)}\Big[2x^2-(a+b+c+d)x\nonumber\\
&+&ab+cd-(d-b)(c-a)\frac{{\bf E}(m)}{{\bf K}(m)}\Big] \ .
\end{eqnarray}
Here
\begin{eqnarray}
G(x)&=&\sqrt{(x-a)(x-b)(x-c)(x-d)}\nonumber\\
m&=&\frac{(d-c)(b-a)}{(d-b)(c-a)}
\end{eqnarray}
and ${\bf E}$ and ${\bf K}$ are the complete elliptic integrals of
the first and second kind \cite{bluebook}. The stress intensity
factor at any one of the tips is obtained by taking the limit
\begin{equation}
K_i=\lim_{x \to x_i}\sqrt{2\pi(x-x_i)}\sigma_{yy}(x,0), \label{limits}
\end{equation}
where $x_i$ is any one of positions of the tips.

\begin{figure}
\centerline{\includegraphics[width=.47\textwidth]{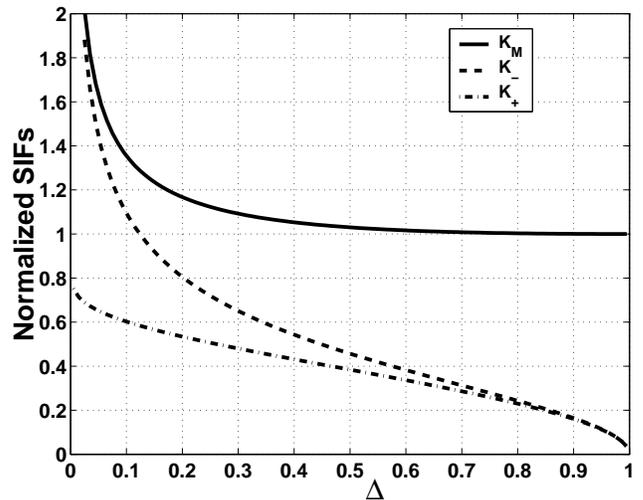}}
\caption{The normalized static stress intensity factors (SIFs) as
a function of $\Delta$. The normalization factor is
$\sigma^{\infty}\sqrt{\pi L/2}$, which is the stress intensity
factor of the macro crack in the absence of the micro crack. We
fixed $\ell_{+}=1$ and varied $\ell_{-}$. Note that the stress
intensity factors obey $K_+<K_-<K_{_{\rm M}}$ and that $K_{_{\rm
M}} \to \sigma^{\infty}\sqrt{\pi L/2}$ as the ratio $\ell/\Delta$
decreases, as expected.} \label{StaticSIF}
\end{figure}

In order to adapt the general configuration to our macro crack and
micro crack configuration we set $a=-L$, $b=0$, $c=\ell_-$,
$d=\ell_+$. Taking the limits in Eq. (\ref{limits}), under the
assumption $L\gg\ell$, we can extract the stress intensity factors
at the three tips
\begin{eqnarray}
K_{_{\rm M}} &=& \sigma^{\infty} \sqrt{\frac{\pi L}{2}}
\sqrt{\frac{\ell_+}{\ell_-}} ~\frac{{\bf
E}(1-\ell_-/\ell_+)}{{\bf K}(1-\ell_-/\ell_+)}\label{SIFs}\\
K_- &=& \sigma^{\infty} \sqrt{\frac{\pi L}{2}} \left[\frac{\ell_+}{\ell_-} ~\frac{{\bf
E}(1-\ell_-/\ell_+)}{{\bf K}(1-\ell_-/\ell_+)}\frac{1}{\sqrt{\ell_+/\ell_- -1}} \right]\nonumber\\
K_+ &=& \sigma^{\infty} \sqrt{\frac{\pi L}{2}} \left[\Big(1-\frac{{\bf
E}(1-\ell_-/\ell_+)}{{\bf K}(1-\ell_-/\ell_+)}\Big)\frac{1}{\sqrt{1- \ell_-/\ell_+}} \right]
\ . \nonumber
\end{eqnarray}
Note that the common pre-factor $\sigma^{\infty} \sqrt{\pi L/2}$
is just the stress intensity factor of the macro crack in the
absence of the micro crack and serves here as the scale of the
three stress intensity factors. In Fig. \ref{StaticSIF} we present
the three stress intensity factors as a function of $\Delta$. In this
example we kept the macro tip at $L$ and the right micro tip at
$\ell_+$ fixed while $\ell$ was changed. We note that the stress
intensity factor of the macro crack goes to the single crack
result (unity in the reduced coordinates of Fig. \ref{StaticSIF})
when $\ell/\Delta \to 0$. Similarly, the stress intensity factor
at $\ell_+$ goes to unity when $\Delta\to 0$, since also in that
limit we remain with one crack. This last fact is not easily seen
in Fig. \ref{StaticSIF} since the upturn towards unity is very
rapid, occurring just before coalescence.

\subsection{The dynamic problem}

\begin{figure}
\centerline{\includegraphics[width=.47\textwidth]{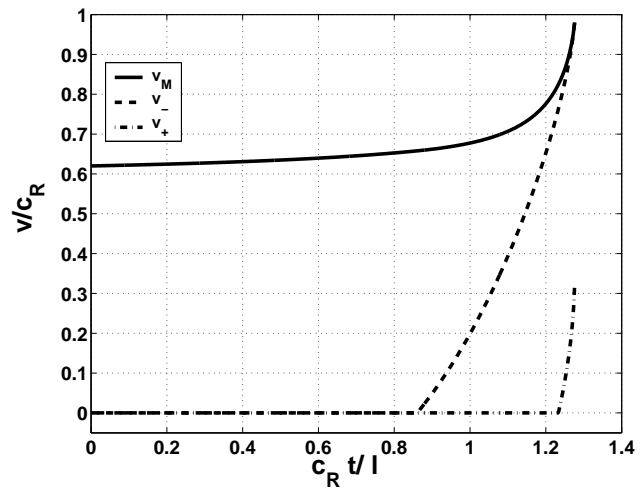}}
\caption{The three crack tip velocities for an interaction event
when a macro crack travelling at an initial velocity $v_{_{\rm
M}}=0.62c_R$ interacts with a colinear micro crack of length $\ell
= 5$ positioned at $\Delta = 5$. The figure shows the normalized
velocities $v/c_R$ as a function of the normalized time $c_R
t/\ell$.} \label{event}
\end{figure}

Using the static stress intensity factors Eqs. (\ref{SIFs}) in Eqs. (\ref{DynSys}) we can solve
numerically for the dynamics of the three tip positions. An example of the ensuing
dynamics is exhibited in Fig. \ref{event}.
We note that what is seen in this picture is typical to all the
conditions that we have considered: the macro crack is first
accelerated, then the left tip of the micro crack meets the
fracture criterion Eq. (\ref{crit}) and accelerates towards the
macro crack; after some time lag, the right tip meets the fracture
criterion and starts to move and attains at coalescence a lower
velocity than the original macro crack.

To connect to velocity fluctuations observed in experiment we
reinterpret the data in Fig. \ref{event} as it would be seen by an
observer. We are physically motivated by the fact that as a result
of a finite measurement resolution, below a critical separation
$\Delta_c$ the macro crack tip and the outer tip of the micro
crack are indistinguishable and the measured velocity is a result
of some averaging. Therefore, let us define the {\em experimental}
velocity $v_{exp}$ as
\begin{equation}
\label{expV}
v_{exp} \equiv \left\{\begin{array}{ll}
v_{_{\rm M}} &\quad\text{for}\ \ \Delta > \Delta_c \\
               (v_{_{\rm M}}+~~v_+)/2 &\quad\text{for} \ \ \Delta < \Delta_c
\end{array}\right.
\end{equation}
\begin{figure}
\centerline{\includegraphics[width=.48\textwidth]{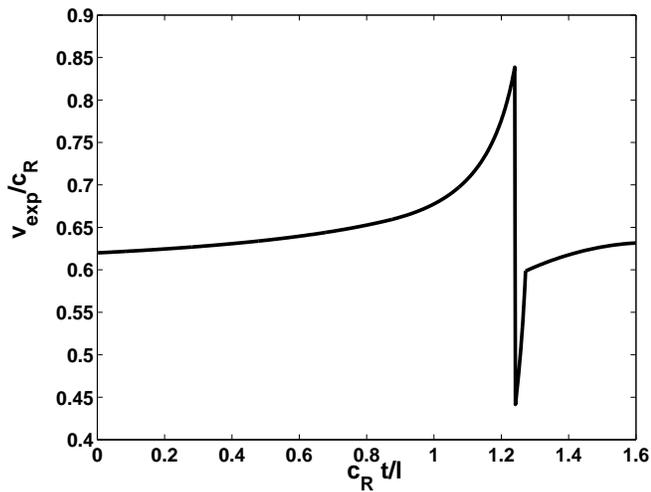}}
\caption{The normalized experimental velocity $v_{exp}/c_R$ as a
function of the normalized time $c_R t/\ell$. The experimental
velocity was calculated according to Eq. (\ref{expV}) with
$\Delta_c \ll \ell$.} \label{shape}
\end{figure}
Figure \ref{shape} shows the experimental velocity $v_{exp}$
during an interaction event. The rise in velocity after the steep decline
is the second of Eqs. (\ref{expV}). The last branch in which the velocity returns
to the pre-collision value is out of the scope of the present model and had
been added by hand for the sake of illustration. The point to stress is that the dynamic
interaction event generates a typical large and rapid velocity
``fluctuation". It should be noted that we do not consider here
the effect of the nucleation of the micro crack on the macro crack
velocity. Physically we expect this effect to produce a sudden
deceleration of the macro crack prior to the effect shown in Fig.
\ref{shape}; this is expected since the energy supply to the crack
tip region should be partitioned between the nucleation process of
the micro crack and the fracture process of the macro crack. This
effect will be taken into account in future work where our current
model will be coupled to a reasonable nucleation theory.

\section{Simulational support}
\label{simul}

Since our reduction to ordinary differential equations rests on
the {\em assumption} of the product structure (\ref{approx}) for
the dynamic stress intensity factors, we must test the quality of
the approximation by numerical simulations. We employ lattice
simulations as described bellow.
\subsection{Lattice simulations}
Lattice models
\cite{slepyan-mode3,slepyan-mode1,marder1,marder2,kess-lev,kess,kess-lev-pech}
provide a convenient, concrete, and physically sensible method of
realizing crack dynamics.  The material is represented by a
lattice of mass points connected by Hookean springs.  Fracture is
achieved when a spring exceeds a certain critical extension.  In
certain special cases, analytic solutions for static and steadily
moving cracks can be obtained.  In general, the model can be
easily simulated.  A major advantage of this class of models is
that the process zone is quite small, on the order of a few
lattice spacings.  Thus, already on scales of 50 or so lattice
spacings, continuum dynamics is very well realized. It has been
recently demonstrated \cite{kess-lev-univ} that the universality
assumption underlying linear elastic fracture mechanics, namely
that the instantaneous crack velocity is only a function of the
stress intensity factor at that moment, is extremely well
satisfied by the lattice dynamics.

For our present purposes, we use the machinery developed in
\cite{kess-lev-univ}.  We work with a square lattice with nearest-
and next-nearest- neighbor bonds, with the ratio of the spring
constants chosen to give isotropic elasticity. As we are
interested in cracks that propagate along the midline, we allow
only bonds that cross the midline to break. We start with a
lattice under fixed-displacement loading at the top and bottom,
with bonds broken according to the desired initial configuration,
Fig. \ref{Sketch}.  We relax this lattice using a multi-grid
approach to accelerate convergence.  At this point, we manually
break the bond at the end of the macro crack, and monitor the
subsequent sequence of bond breakings.

\subsection{Results of lattice simulations}

\begin{figure}
\centerline{\includegraphics[width=.49\textwidth]{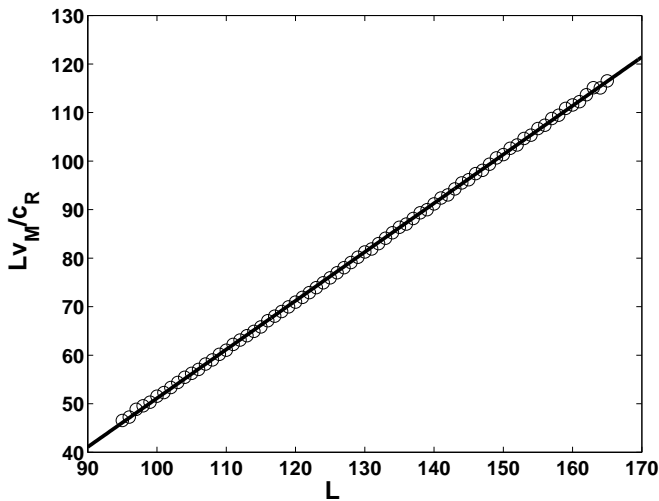}}
\caption{The simulation data of $Lv_{_{\rm M}}/c_R$ as a function
of $L$ is shown in circles and the linear fit for this data is
shown in solid line. It is clear that the functional form
suggested by Eq. (\ref{fit}) indeed describes the single crack
dynamics in the lattice simulation. The fit is consistent with the
assumptions since the slope $\alpha$ is very close to one.}
\label{LinearFit}
\end{figure}

In order to test the quality the product structure approximation
we should first show that the analytic equation (\ref{BasicEq})
for the dynamics of a {\em single} macro crack describes correctly
the corresponding dynamics in the lattice simulations. Multiplying
Eq. (\ref{BasicEq}) by $L$ we obtain
\begin{equation}
L \frac{v_{_{\rm M}}}{c_R} = \alpha L + \beta \ , \label{fit}
\end{equation}
where $\alpha$ is predicted to equal one. The parameter $\beta$
relates the material properties and boundary conditions of the
lattice experiment to those in the continuum model. We simulated a
single crack propagating without any additional damage, and
measured $v_{_{\rm M}}/c_R$ as a function of $L$. Next we fitted
the data to Eq. (\ref{fit}). Figure \ref{LinearFit} shows that the
functional form given by Eq. (\ref{fit}) describes well the single
crack dynamics in the lattice simulation. Moreover, we found that
$\alpha \approx 1$; thus our procedure appears internally
consistent. It should be noted that, notwithstanding the excellent
agreement evidenced in Figure \ref{LinearFit}, there are a number
of uncontrolled approximations at play here.  First, the
functional form in Eq. (\ref{fit}) is based on the assumption of a
constant fracture energy.  In fact, the fracture energy for our
theoretical ``lattice'' material has been calculated, and it is
not constant. Second, the simulation employs constant displacement
boundary conditions in a finite strip (though, in order to mimic
infinite medium, we specialized for times that do not allow wave
interactions with the outer boundaries), and the theory assumes
fixed stress at infinity. The excellence of the fit despite all
this is somewhat unexpected, and bears further study.

\begin{figure}
\centerline{\includegraphics[width=.49\textwidth]{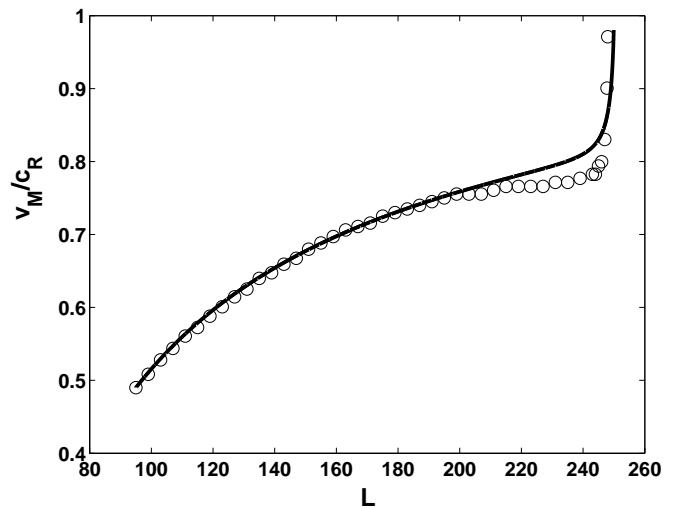}}
\caption{The simulation data of $v_{_{\rm M}}/c_R$ as a function
of $L$ is shown in circles and the analytic approximation
predictions are shown in solid line. In this simulation we used $L
\approx 245$ (the length of the macro crack in the interaction
region) and $\ell=10$. The analytic approximation predicted the
simulated data up to an error of $6.5\%$, even though the ratio
$L/\ell$ was not very large.} \label{comparison}
\end{figure}

To directly test our product structure approximation Eqs.
(\ref{approx}) we used the value of $\beta$ obtained by the linear
fit and performed lattice simulation where a macro crack interacts
with a colinear micro crack. We encounter a difficulty in the
simulations since the micro crack tips were trapped even when the
stress intensity factor exceeds the material threshold
(\ref{crit}). This known phenomenon of lattice trapping is an
artifact of the lattice structure; to overcome it we fixed the
micro crack tips also in the analytic calculation. An example of
the comparison between the simulation data and the analytic
approximation is shown in Fig. \ref{comparison}. We found that our
analytic approximation agrees with the simulated data, with
deviations that are typically small. The largest errors are
smaller than about $6-7\%$ even for $L/\ell \approx 25$. For
larger ratios, which is the expected physical regime, we expect
better approximations. Note that the fact that the analytic
approximation overestimates the velocity of the macro crack is
expected on physical grounds. Our product structure approximation
relies on the fact that for short distances the information on the
positions of the cracks tips, carried by elastic waves, flows
almost instantly even if the typical cracks propagation velocities
are of the order of $c_R$. In reality it takes finite time for the
stresses to reorganize themselves according to the new cracks tips
positions and generally the energy release rate is lower than in
our approximation.

The main conclusion of this section is that the product structure
approximation Eqs. (\ref{approx}) gives very good predictions for
the model problem studied here. We propose to interpret this in a
broader way, and to test the applicability of this approximation
in other contexts.

\section{Summary and conclusions}

This study has two aims; on the one hand, we are interested in the
velocity fluctuations seen in dynamical crack propagation, and
proposed here that the linear-elastodynamics interactions with
micro cracks may very well be responsible for them. We did not
address the reasons for the existence of the micro cracks - these
may be there {\em ab-initio} or get born by the high stresses near
the tip of the advancing cracks. On the other hand, we are after
simplified methods of analysis of crack propagation in non-trivial
environments. The technical difficulty of solving the full
dynamical equations calls for approximate methods that work. With
the motivation presented in Sect. \ref{Approx} we demonstrated how
the assumption of product structure for the stress intensity
factors reduces the dynamics to a set of ordinary differential
equations that are easily solved. The gratifying agreement with
the lattice simulations emboldens us to propose this as an
approach that may find applications in other contexts of interest.
Only future work will help to strengthen or delineate the
usefulness of this approach.

To further connect the interaction model to experimental
observations one should couple our theory to a physically
motivated model that will determined the conditions for micro
cracks nucleation. Such a model will potentially predict the
appearance of distributed damage in the process zone and the
interaction with the macro crack may be related to the roughness
of crack surfaces and the quasi-periodicity of the velocity
fluctuations.

\label{summary}


\begin{thebibliography}{99}

\bibitem{84R-CK_1}
 K. Ravi-Chandar and W.G. Knauss, Int. J. Frac. \bf{26}\rm, 65 (1984).

\bibitem{84R-CK_2}
 K. Ravi-Chandar and W.G. Knauss, Int. J. Frac. \bf{26}\rm, 141 (1984).

\bibitem{95SF}
 E. Sharon, S.P. Gross and J. Fineberg, Phys. Rev. Lett. \bf{74}\rm, 5096 (1995).

\bibitem{96SF}
 E. Sharon and J. Fineberg, Phys. Rev. B \bf{54}\rm, 7128 (1996).

\bibitem{97R-CY}
 K. Ravi-Chandar and B. Yang, J. Mech. Phys. Solids \bf{45}\rm, 535 (1997).

\bibitem{98R}
 K. Ravi-Chandar, Int. J. Frac. \bf{90}\rm, 83 (1998).

\bibitem{99SF}
E. Sharon and J. Fineberg, Nature {\bf 333}, 397 (1999).

\bibitem{99FM}
 J. Fineberg and M. Marder, Phys. Rep. {\bf 313}, 1 (1999).

 \bibitem{98F}
L. B. Freund, {\em Dynamic Fracture Mechanics}, (Cambridge, 1998).

\bibitem{72F}
 L.B. Freund, J. Mech. Phys. Solids {\bf 20}, 141 (1972).

\bibitem{75Kos}
 B.V. Kostrov, Int. J. Frac. {\bf 11}, 47 (1975).

\bibitem{03A-B}
 M. Adda-Bedia, Preprint, (2003).

\bibitem{99B}
K. B.  Broberg, {\em Cracks and Fracture}, (Academic Press, 1999).

\bibitem{bluebook}
M. Abramowitz and I.A. Stegun, {\em Handbook of Mathematical
Functions} (Dover, 1975).

\bibitem{slepyan-mode3}L. I. Slepyan, Sov. Phys. Dokl. {\bf 26} 538 (1981).

\bibitem{slepyan-mode1}
Sh. A. Kulamekhmetova, V. A. Saraikin
and L. I. Slepyan, Mech. Solids {\bf 19}, 102 (1984).

\bibitem{marder1}
M. Marder and X. Liu, \prl {\bf 71}, 2417 (1993).

\bibitem{marder2}
M. Marder and S. Gross, J. Mech. Phys. Solids {\bf 43}, 1 (1995).

\bibitem{kess-lev}
D. A. Kessler and H. Levine, \pre {bf 59}, 5154 (1998).

\bibitem{kess}
D. A. Kessler, \pre {bf 61}, 2348 (2000).

\bibitem{kess-lev-pech}
L. Pechenik, H. Levine andD. A. Kessler, J. Mech. Phys.
Solids {\bf 50}, 583 (2002).

\bibitem{kess-lev-univ}
D. A. Kessler and H. Levine, \pre {\bf 68}, art. no. 036118 (2003).


\end{thebibliography}
\end{document}